\documentstyle[]{article}
\renewcommand{\baselinestretch}{1.2}
\begin{document}
%
%
\title{Relational Approach to Spin Networks 
      \footnote{Partly based on the author's contribution to 69th Annual 
       Meeting of the German Physical Society: Physics Since Einstein, 
       Berlin, Germany, 4-9 Mar 2005. \cite{ws1}}}
\author{Walter Smilga \\ Isardamm 135 d, D-82538 Geretsried, Germany \\
e-mail: wsmilga@compuserve.com}
\maketitle 
\renewcommand{\baselinestretch}{1.2}

\begin{abstract}
Individual spinors in a SU(2) spin network are described by their 
relations to the background spin network. 
A 'covariant' formulation of these relations yields the de Sitter group
SO(3,2) as the fundamental symmetry group.
Locally this symmetry group is approximated by the Poincar\'e group, 
which leaves invariant (certain) clusters of spinors. 
The calculated masses of these clusters reproduce the lepton spectrum.  
Corrections to the approximate Poincar\'e group, based on the 
exact SO(3,2) symmetry, deliver interaction terms, identical to those of 
the standard model. 
In addition, gravitation is obtained.
The calculation of the fine-structure constant reproduces Wyler's formula. 

\end{abstract}


\renewcommand{\baselinestretch}{1.0}

\section{Introduction}

The literature on spin networks is dominated by their graph theoretical 
aspects. 
In this article the notion of spin networks will be used in an 
elementary sense. It will denote a set of two-component spinor states, that 
can be linked together (and unlinked again) according to the quantum 
mechanical rules for coupling angular momentum.

The word ``Relational" in the title refers to the concept of 
\begin{it}Relational Quantum Mechanics\end{it}, which was 
developed by C. Rovelli\cite{cr} and others.
The approach to spin networks, presented here, will essentially study
relations between individual spinors and a background network.

The approach will also illustrate the role of reduction and emergence
in particle physics. 
It may, therefore, be considered as a contribution to the dispute about 
reductionism versus emergence in physics research that recently has been 
brought up again \cite{rl}.

There is a well-known dictum of Bohr's:
\begin{quote}
``Physics is not about reality, but about what we can say 
about reality."\cite{bh}
\end{quote}
To those, who have some background of information theory, this dictum 
suggests that the object of physics is information, information about reality.
Particle physics, in particular, is then the part of physics that is in
search for the most elementary building blocks, not of reality, but of 
information about reality.

The question, whether there may or may not exist a one-to-one relation 
between information and ``reality", is not relevant for 
the following.
This question cannot be answered by physical means.
Therefore, it is not a subject of particle physics, but rather, if at all, 
of metaphysics.
For the following it will be sufficient to understand ``reality" simply as 
the collection of all available information about reality.

The relevance of information theory for understanding quantum mechanics 
has been accentuated especially by John Wheeler \cite{jaw}, who coined the 
phrase ``It from Bit".
In fact, the decomposition of information into smaller and smaller pieces 
finally results in a set of yes-no decisions, or binary elements, or, simply,
bits.
Hence, the question for the most elementary building blocks of information
about reality is very easily answered:
They are neither strings nor branes, but simply binary elements.
Therefore, the ultimate goal of every ``reductionist" should be the 
formulation of physics in terms of binary elements.

Many of us expect to find the most elementary building blocks of reality 
at scales of the Planck length.
At these scales it would hardly be possible to experimentally identify 
binary elements.
But, assuming that there are basic binary elements at accessible scales, 
are we really prepared to observe such elements? 
Would we even realize that it is a binary element, when we 
encounter it in, say, a scattering experiment?
Can binary elements at all turn up as isolated physical entities?
In case they can, what would binary elements look like?
To answer these and other questions we must, obviously, develop a theoretical
understanding of how single binary elements would behave, if they were 
observed in a suitably designed experimental setup.

When we accept binary elements as the ultimate building blocks of reality,
we are faced with the challenge to make understandable, how 
such elements of particle physics, as space-time, masses, charges 
and interaction can be derived from binary elements.
In face of the tremendous difficulties of string theory, to reach its
goals on the basis of strings and branes, the intention to base physics 
on even more elementary building blocks, may appear highly adventurous.
Nevertheless, in the past, several serious authors have attempted to 
establish connections between sets of binary elements and fundamental
structures of physics.

The quantum mechanical generalisation of binary elements are spinors.
These are described by states in a SU(2)-symmetric Hilbert space. 
A set of spinors then forms a spin network in the elementary sense defined
above.
We will see below that, in principle, it makes no difference, whether we start
from a set of binary elements or from a set of spinors.
 
Spin networks were introduced by Roger Penrose \cite{rp} almost 40 years ago, 
in an attempt to describe the geometry of space-time in a purely combinatorial 
way. 
Penrose studied systems of two-component spinors, which represent the simplest 
quantum mechanical objects. 
He was able to show that large systems of such spinors, generate properties of 
angular directions in three-dimensional space. 
Despite this success, the concept of SU(2) based spin networks was considered 
not rich enough to describe distances \cite{rpwr}. 

Another attempt to derive space-time from ``binary alternatives" was 
undertaken by C. F. von Weizs\"acker \cite{cfw1,cfw2} and co-workers. 
In this concept a physical particle is considered as an aggregate of
$10^{40}$ binary alternatives, representing the knowledge about the particle.
Although indications for a Poincar\'{e} symmetric space-time structure were 
obtained, the work remained at an experimental stage without having 
established a definite link to empirical particle physics.
 
Spin networks were re-discovered by Rovelli and Smolin as a basis for
studies on quantum gravity \cite{crls1,crls2}.
In quantum gravity spin networks are employed to build models of space-time 
at Planck scales.
In contrast to these models, the scales encountered in the present article 
will turn out to be experimentally accessible.

The relational approach to spin networks is essentially guided by Bohr's 
statement\cite{mj} 
\begin{quote} 
``Physical phenomena are observed \begin{em}relative\end{em} to different 
experimental setups" (accentuation by the author).
\end{quote}
Bohr's dictum in a concise way states that we do physics in a
``relational" way.
In the context of spin networks, Bohr's statement can be understood as an 
invitation, to establish a mathematical description of spinors in a spin
network, by their relations to other subsets of the spin network. 
These subsets then stand for Bohr's experimental setups.
We will find that such relations, in fact, provide individual spinors with 
``physical" properties that can be compared to well-known elements of 
the standard model. 
Therefore, this paper can be understood as an attempt to base the standard
model on spin networks.

The outline of the paper is as follows:
In section 2 we will construct from spinors a quasi-classical 
reference object, which can be used as a simple model of an experimental 
setup.
We will obtain the de Sitter group SO(3,2) as symmetry group of the state 
space of this reference object.
A local approximation to this symmetry delivers the Poincar\'e group
as the local kinematical group of the reference.
In section 3, four-component Dirac spinors are introduced as means to 
describe the coupling of two-component spinors to a reference.
In section 4, a covariant formulation of this coupling results in a 
description of Dirac spinors in space-time, by solutions of Dirac's 
equation.  
Sections 5 and 6 describe, how mass relations for different types of 
linking spinors to references reproduce the leptonic mass spectrum.
In Section 7 the interaction term of quantum electrodynamics is derived
as a SO(3,2) correction to the approximate Poincar\'e group.
In section 8, the electromagnetic coupling constant is calculated. 
The result is Wyler's formula for the fine-structure constant.
Sections 9 and 10 shortly discuss how the interaction terms of quantum 
chromodynamics and weak interaction result from SO(3,2) corrections.
Section 11 shows that gravitation is a natural result of the basic
SO(3,2) symmetry. 
Section 12 identifies spin networks are the relational form of sets of 
binary elements.
Section 13 concludes with the statement that binary elements neither 
need nor allow for a deeper explanatory base.

\section{Quasi-classical description of large subsets}

Following Bohr's dictum we will develop a mathematical theory of individual 
spinors \begin{em}relative\end{em} to certain reference subsets of the spin 
network.
We would be well advised to choose the references in a most general, but 
nevertheless realistic way.
A suitable candidate for a reference object is the counterpart to a classical 
rigid body, capable of moving in space-time and rotating with respect to three 
spatial axes. 
Such an object may carry a coordinate system.
Then the kinematical states of the reference object can be characterised by 
all possible coordinate transformations.
Of course, we want the theory to be valid for all states of the reference 
object.
This means, the theory shall be covariant with respect to a group of 
coordinate transformations that act transitive on the state space of the 
reference object.
Therefore, we have two things to do:
Firstly, construct a reference object from spinors that comes close to a 
classical body.
Secondly, determine the transformation group that describes the kinematics of 
this object.

We construct a reference object from a set of spinors by linking them together 
according to the quantum mechanical rules for coupling angular momenta.
Assume that the spinors are linked to total states with well-defined SU(2) 
quantum numbers.
If the reference is sufficiently large, it can be treated by the limit of
large quantum numbers, which means quasi-classically. 
Using the three infinitesimal generators $j_{12}, j_{23}, j_{31}$ of SU(2), 
as defined below, we can rotate the reference with respect to three orthogonal 
axes in a quasi-continuous way.
These rotations form the group SO(3).
For a given total angular momentum, however, the axis of rotation is fixed,
as a result of conservation of angular momentum. 
Therefore the (quasi-classical) motion of the reference will essentially be 
restricted to a 2-dimensional plane.
This means that the reference as a whole performs an ``orbital" rotational 
movement with respect to a (distant) centre in a quasi-continuous way.
Hence, a first contribution to the kinematical group is the group SO(2).

Assume that we divide the reference object into several parts. 
This allows us to take differences of individual angular momenta. 
To reduce the number of degrees-of-freedom, let us assume that the parts
are fixed relative to each other, so that they form the equivalent of a
rigid body.
Then the operators of differences of angular momentum can be used as 
infinitesimal generators to rotate the object as a whole relative to its 
(local) centre.
These rotations determine the ``orientation" of the object relative to
three orthogonal axes.
The rotations again form a group SO(3), which extends the kinematical
group to the product group SO(3)$\times$SO(2).

Assume now that a naive observer is moving together with the reference object.
Then he will not immediately take notice of the orbital motion.
However, he may observe that other objects are moving relative to his own 
coordinate system.
He will inevitably develop a perception of ``time", which ``passes" 
although his own coordinate system does not seem to move. 
Probably he will then decide to measure the elapse of time by comparing it 
with some suitable periodic motions of other objects. 
Therefore, the SO(2)-part of the kinematical group shows properties of 
``time", whereas the SO(3)-part describes the orientation of the rigid body 
in 3-dimensional ``space". 

In a next step the observer may develop techniques to move his position 
relative to the 3-dimensional coordinate system. 
Alternatively, he may move the reference object together with its coordinate 
system relative to his own position.
Such actions are described by boost operations, which are coordinate 
transformations that let the SO(3)-coordinate system move in the direction 
of a coordinate of the SO(2)-system.
The boost operations extend, as will be shown below, the kinematical group to 
SO(3,2).
This group contains all coordinate transformations that our hypothetical 
observer is able to carry out, or, alternatively, the reference object may be
subjected to.
A description of a single spinor that remains valid after one of these 
transformations, must therefore be covariant with respect to the operations 
of SO(3,2).
In this way, SO(3,2) becomes the basic symmetry group of a theory of spin 
networks that will be developed in the following.

To examine the kinematical group in more detail, let us represent the 
infinitesimal generators of the group operations of the SO(3)-part by 
differential operators. 
These will act on functions in an auxiliary 3-dimensional 
parameter space spanned by the coordinates $x_1, x_2, x_3$.
Later it will become clear that we can form states that are ``localized"
at any given ``point" of this parameter space, thus making it a 
``physical space" in the usual sense.

Rotations of the coordinate system of the reference are then generated by 
three well-known differential operators
\begin{equation}
j_{ij} = i( x_i\,\partial_j - x_j\,\partial_i),\;\;\; i,j=1,2,3\;.\label{1-3}
\end{equation}

The SO(2)-part of SO(3,2) can be represented by rotations in a 2-dimensional
parameter space spanned by two additional parameters $x_0$ and $x_4$.
Then the infinitesimal generator of SO(2) rotations is given by
\begin{equation}
j_{04} = i(x_0\, \partial_4 - x_4\, \partial_0) \; .              \label{1-4}            
\end{equation}

Both parameter spaces, spanned by coordinates $(x_1, x_2, x_3)$ and 
$(x_0, x_4)$, respectively, are Euclidean.
We can combine them to a 5-dimensional pseudo-Euclidean space with the metric 
\begin{equation}
g_{ab} = diag(\;+1,-1,-1,-1,+1\;), \;\;a,b = 0,...,4\;.  \label{1-5}
\end{equation}
This allows us to express the boost operations by pseudo-rotations in the 
$(0,i)$-plane, $i = 1,2,3$, as known from Lorentz transformations. 
The boost operations are then generated by 
\begin{equation}
j_{0i} = i(x_0\, \partial_i - x_i\, \partial_0),\;\;\; i=1,2,3 \;.\label{1-6}
\end{equation}
The proof is by verifying the commutation relations of the homogeneous 
Lorentz group.
There is a second set of pseudo-rotations, which are generated by 
\begin{equation}
j_{\mu4} = i(x_\mu\, \partial_4 - x_4\, \partial_\mu),\;\;\; \mu = 0,1,2,3 \;. 
\label{1-7}
\end{equation}

It can easily be verified that the differential operators 
(\ref{1-3},\ref{1-4},\ref{1-6},\ref{1-7}) satisfy the commutation relations 
of SO(3,2):
\begin{equation}
[j_{\mu\nu}, j_{\rho\sigma}] = 
-i[g_{\mu\rho} j_{\nu\sigma} - g_{\mu\sigma} 
j_{\nu\rho} + g_{\nu\sigma} j_{\mu\rho} 
- g_{\nu\rho} j_{\mu\sigma}] \mbox{ , }                         \label{1-10}
\end{equation}
\begin{equation}
[j_{\mu4}, j_{\nu4}] = -i j_{\mu\nu} \mbox{ , }                 \label{1-11}
\end{equation}
\begin{equation}
[j_{\mu\nu}, j_{\rho4}] 
= i[g_{\nu\rho} j_{\mu4} - g_{\mu\rho} j_{\nu4}] \;.            \label{1-12}	 
\end{equation} 
 
Now consider operators (\ref{1-7}) in the neighbourhood of the point
${\mathcal{O}} =(x_\mu=0, x_4=R)$.
For $R$ large compared to the components $x_\mu$ the operator $j_{\mu4}/R$
can be approximated by 
\begin{equation}
p_\mu = -i \, \partial_\mu,  \;\;\; \mu = 0,1,2,3 \;. \label{1-8}
\end{equation}
The operators $p_\mu$ commute and act as generators of translations
in the 4-dimensional subspace defined by $(x_0, ... , x_3)$.
Together with (\ref{1-3}) and (\ref{1-6}) they generate the 
inhomogeneous Lorentz group (Poincar\'e group) P(3,1).
This approximation, valid in the neighbourhood of a given point of the 
parameter space, is known as group contraction\cite{iw,fg}.

From eigenstates of the three spatial translation operators 
we can, by superposition, construct states that are localized at any
position $x_i, i = 1,2,3$, in the neighbourhood of ${\mathcal{O}}$.
The time-like translation operator then describes the evolution of
this state in time.
The translation operators therefore give the abstract parameters
$(x_0, ... , x_3)$ the familiar meaning of space-time.

Obviously $x_i$ and $p_k$ satisfy the well-known 
commutation relations
\begin{equation}
[x_i, p_i] = i \delta_{ik} \, .
\end{equation}

Without any indication of a ``fourth dimension", the time coordinate is 
derived from the parameterisation of the orbital angular momentum, which 
is subject to a SO(3) symmetry. 

That ends the construction of a reference object from a subset of the
spin network.
Its kinematics can be expressed by coordinate transformations within the 
parameter space.
The transformations form the group SO(3,2).
The reference object can be treated quasi-classically, but nevertheless it
is still a quantum mechanical object.

\section{Spinors within spin networks}

In this section a quantum mechanical description of a single spinor relative 
to a quasi-classical reference object will be given.

Spinor states form a complex vector space built from two-component base vectors
\begin{equation}
|u\rangle =  \left( \begin{array}{*{1}{c}} 1 \\ 0 \end{array} \right)  
 \;\; \mbox{ and } \;\;
|d\rangle = \left( \begin{array}{*{1}{c}} 0 \\ 1 \end{array} \right)\;.  
                                                                \label{2-1}
\end{equation}
It is well-known that the Pauli matrices
\begin{equation}
\sigma_1 = \left( \begin{array}{*{2}{c}} 0 & 1 \\ 1 & 0 \end{array} \right),                       
 \;\;\;\;
\sigma_2 = \left( \begin{array}{*{2}{c}} 0 & -i \\ i & 0 \end{array} \right),
 \;\;\;\;
\sigma_3 = \left( \begin{array}{*{2}{c}} 1 & 0 \\ 0 & -1 \end{array} \right).
                                                                \label{2-4}  
\end{equation}       
are the generators of SU(2) transformations within this vector space.  
 
Now consider a reference object with well-defined quantum numbers 
$(S, S_3)$ with respect to SU(2). 
By adding another single spinor to this state, we can obtain either a state
\begin{equation}
(S+\frac{1}{2}, S_3+\frac{1}{2})     \label{3-1a}
\end{equation}
or 
\begin{equation}
(S-\frac{1}{2}, S_3-\frac{1}{2})     \label{3-1b}
\end{equation}
or any linear combination of these states.
Which of these states is obtained, depends on the orientation of the
single spinor relative to the reference state. 
The relative orientation is neither a property of the single spinor nor of 
the reference state, but rather a property of the combined system.
However, if we keep the orientation of the reference fixed, we are 
able to formally express the two ways of coupling as a property of the spinor 
itself. 
We can do this by simply doubling the spinor components in the following way:

\begin{eqnarray}
|u_1\rangle &=& \left(\begin{array}{*{1}{c}} 1 \\ 0 \\ 0 \\ 0 \end{array}\right)  
 \; \mbox{ , } \;
|u_2\rangle  =  \left(\begin{array}{*{1}{c}} 0 \\ 1 \\ 0 \\ 0 \end{array}\right)  
                                                               \label{3-2a}\\
\nonumber \\
|v_1\rangle &=& \left(\begin{array}{*{1}{c}} 0 \\ 0 \\ 0 \\ 1 \end{array}\right)  
 \; \mbox{ , } \;
|v_2\rangle  =  \left(\begin{array}{*{1}{c}} 0 \\ 0 \\ 1 \\ 0 \end{array}\right).  
                                                                \label{3-2b}
\end{eqnarray}

The first group (\ref{3-2a}) of base states describes a spinor coupled to the 
reference object in the sense of (\ref{3-1a}).
The second group (\ref{3-2b}) describes a spinor coupled to the 
reference object in the sense of (\ref{3-1b}).
The matrix
\begin{equation}
\gamma^0 = 
\left( \begin{array}{*{2}{c}} I & 0 \\ 0 & -I \end{array} \right), 
                                                                \label{3-4}
\end{equation}
where $I$ is the $2\times 2$ unit matrix,
then delivers an eigenvalue of $+1$, if applied to the first group of spinors 
(\ref{3-2a}), and $-1$, if applied to the second (\ref{3-2b}).

This formulation is not yet covariant with respect to $SO(3,2)$.
To make it covariant, we have to find $4\times4$ matrices that transform
together with $\gamma^0$ in the same way as the state of the reference is 
transformed by the operations of $SO(3,2)$.
In other words, we have to find a representation of $SO(3,2)$ by $4\times4$
matrices.

The representation of rotations is straightforward.
Their generators are obtained from Pauli matrices in the following form
\begin{equation}
\sigma_{ij} = \epsilon_{ijk} \left(\begin{array}{*{2}{c}} 
\sigma_k & 0 \\ 0 & \sigma_k \end{array}\right),
\; i,j,k = 1,2,3 \;.                                               \label{3-3} 
\end{equation}
Boost operations are generated by the $4\times4$-matrix   
\begin{equation}
\sigma^{0k} = -\sigma^{k0} = \left( \begin{array}{*{2}{c}} 0 & i\sigma_k \\
       -i\sigma_k & 0 \end{array}\right)\;.        \label{3-6a}
\end{equation}

When we close the algebra of the matrices defined so far, 
with respect to their commutation product, we find additional matrices
\begin{equation}
\gamma^k = 
\left( \begin{array}{*{2}{c}} 0 & \sigma_k \\
 -\sigma_k & 0 \end{array}\right) \;.                        \label{3-6}
\end{equation}
 
We can combine the indices $0$ and $k$ to an index $\mu = 0,\ldots,3$, and
use $g_{\mu\nu}=$ diag $(+1,-1,-1,-1)$ in the usual way to raise and lower
indices.

The matrices (\ref{3-4}) and (\ref{3-6}) are Dirac's $\gamma$-matrices in the 
so-called standard or Dirac representation. 
$\gamma$-matrices  satisfy the well-known anti-commutation relations
\begin{equation}
\{\gamma_\mu, \gamma_\nu \} = 2 g_{\mu\nu} \;.                  \label{3-7a}
\end{equation} 
and the commutation relations
\begin{equation}
\frac{i}{2} \, [\gamma_\mu, \gamma_\nu ] = \sigma_{\mu\nu} \;.  \label{3-7b}
\end{equation} 

The $4\times4$-matrices
$s_{\mu\nu}$ and $s_\mu$, built from Dirac matrices, 
\begin{equation}
s_{\mu\nu} :=\, \frac{1}{2} \sigma_{\mu\nu}             
\mbox{   and  }  
s_\mu :=\, \frac{1}{2} \gamma_\mu                            \label{3-9}
\end{equation} 
form a representation of SO(3,2). 
The proof is by verifying the commutation relations 
(\ref{1-10},\ref{1-11},\ref{1-12}) of SO(3,2).

In the following, 4-component spinor states, obtained from the base states
(\ref{3-2a},\ref{3-2b}) by applying transformations generated by the matrices
(\ref{3-9}), will be called Dirac spinors.

To make the vector space of Dirac spinors a Hilbert space, we can add
a scalar product, as known from the treatment of the Dirac equation,
\begin{equation}
\langle \bar{a} | b \rangle  \mbox{   with   } 
\langle \bar{a} | = \langle a | \gamma^0 \; .                 \label{3-8}
\end{equation}

The introduction of Dirac spinors allows to keep track of the coupling
of a single two-component spinor to a reference object. 
Their introduction does not mean any modification of the basic
two-component spin network. 
It only defines an alternative view on the basic spin network.

\section{Spinors in space-time}

In this section we will talk about space-time properties of spinors
in the neighbourhood of the point of contraction $\mathcal{O}$.
Therefore, whenever advantageous, we will replace SO(3,2) by the 
approximate P(3,1).

Let $|P\rangle$ be the state of a quasi-classical reference object 
with momentum $P$. 
By adding a spinor the momentum is changed by an increment $p$ to
\begin{equation}
|P'\rangle = |P + p\rangle \;.                      \label{5-2}                  
\end{equation}   
In the rest frame, we have $P' = (P_0 + p_0, 0, 0, 0)$. 
Since $P_0$ is the approximation to $J_{40}$, the addition of a spinor
increments or decrements the value of $P_0$ by $1/2$. 
Therefore $p_0$ is positive or negative, depending on the coupling of 
the spinor.
This dependency can be expressed by the simple relation
\begin{equation}
(\gamma^0 p_0 - m) |\psi\rangle = 0 , \mbox{ with } m = \mbox{const.}
                                                            \label{5-8}            
\end{equation}
Taking into account the transformation properties of $\gamma^\mu$ with 
respect to SO(3,2) and $p_\mu$ with respect to P(3,1), we can write 
(\ref{5-8}) in a covariant form
\begin{equation}
(\gamma^\mu p_\mu - m) |\psi(p)\rangle = 0 \;.              \label{5-9}            
\end{equation}
This is the Dirac equation with mass $m = 1/2$.
The well-defined value $1/2$ sets a numerical mass scale, which is the 
precondition for the existence of a discrete mass spectrum.

Since the momentum $p_\mu$ in (\ref{5-9}) has been ``borrowed" from the 
reference, the transformation properties of $p$ with respect to the 
Poincar\'{e} group are identical to that of $P$.
But the apparent space-time properties of the spinor do not reflect 
degrees of freedom of the spinor itself, but stand for the 
\begin{em}relation\end{em} of the spinor to a reference object.
There are still only two independent states in the Hilbert space of a spinor.
The parallel or anti-parallel way of coupling to a reference is covered by 
doubling the spinor components to form a Dirac spinor.
The momentum information is coded into the coefficients of the Dirac spinor 
state and reflects the combined state of spinor and reference object.

Consider a momentum eigenstate of the spinor attached to the reference, 
expressed by a plane wave 
\begin{equation}
e^{i(P_\mu + p_\mu)x^\mu} = e^{iP_\mu x^\mu} e^{ip_\mu x^\mu}\;. \label{5-10}
\end{equation}
When we apply a finite translation by a displacement vector $a$ to the spinor 
but not to the reference, this results in a phase factor
\begin{equation}
e^{i p_\mu a^\mu} \;.                                             \label{5-11}
\end{equation}
If at position $a$ relative to the reference there is another reference
object, then this displacement acts as a recoupling of the spinor from
the first reference to the second.
Invariance with respect to P(3,1) ensures conservation of energy-momentum
for the process of recoupling.
Hence, a spinor is able to ``carry" momentum from one reference to
the other.

In this sense the spinor part of (\ref{5-10}) can be regarded as the
description of a particle, capable of moving from one reference to the 
other.
Or as we may say: from one experimental setup to another.

Thus we have arrived at a description of a \begin{em}spinor in 
space-time\end{em}, where the space-time continuum is defined by the
kinematical degrees-of-freedom of macroscopic reference objects. 

The ``position" of the spinor has been defined relative to the position 
of the reference by an offset vector $a$.
Of course, there are other ways to determine the position of an object 
in space.
For example, a position can be defined by the intersection of three
perpendicular planes. 
Each plane can be determined by its distance from one of three
reference objects.
In such a configuration three reference objects contribute to the position
of the spinor.
The determination of the effective mass of a spinor within such a 
configuration is not as trivial as in the case of only one reference. 
In the following section we will derive a ``mass formula", which relates 
the mass of the one-reference case to effective masses in configurations 
with two and three reference objects.

Adding a fourth reference to define a spinor position would lead to an 
over-determined system.
Therefore, in the following we consider only up to three references.

\section{Mass relations}

A few years ago G.~Gonz\'alez-Mart\'in \cite{ggm1,ggm} (G-M in the following)
obtained mass relations, based on an universal structure group SL(4,R).
G-M's idea was that the structure group describes a ``substrate", from which
particles are generated as ``excitations" with certain symmetric and
topological properties, which are associated with subgroups of the structure
group.
G-M found a mass formula for the three massive leptons
\begin{equation}
m_n = 4\pi \left(\frac{16 \pi}{3}\right)^n \; m_e 
\hspace{1cm}  n = 1, 2 \;\; ,                                \label{6-8}
\end{equation}
where $m_e$ is the electron mass and $m_1$ stands for 
the myon mass, $m_2$ to the tauon mass.
With the experimental electron mass of $0.5109989$ MeV, G-M obtained
$m_\mu = 107,5916$ MeV and $m_\tau = 1770,3$ MeV.
(The experimental values are $105,658$ and $1776,99$.)

We will see that our approach to space-time properties of spinors
leads to an explanation of G-M's mass formula on the basis of spin
networks.

Consider three reference objects with entangled wave functions and
a single spinor linked to their wave function.  
Then, instead of (\ref{5-10}), we have a state built from entangled 
contributions of
\begin{equation}
e^{i(P'_\mu + P''_\mu + P'''_\mu + p_\mu)\, x^\mu} . \label{6-0}
\end{equation}
The spinor is now added to three reference objects at the same time.
What is then the effective spinor mass in such a configuration
compared to the single reference situation?

The states of the reference objects belong to a product representation
of SO(3,2). 
To obtain mass relations, the product representation must be made comparable 
to the single reference situation treated in the last section. 
A suitable way of achieving this is to convert the product representation
into a direct sum (or rather integral) of irreducible representations of the 
Poincar\'e group.
Each representation will then contribute the same value of $1/2$ to the mass.
The effective spinor mass is subsequently determined by a sum over all
contributing representations.

In a first step we decompose a quasi-classical representation of SO(3,2) 
into a set of SO(3,1) representations.
Let $S$ denote the group of SO(3,2) transformations. 
Let $L$ denote the subgroup of Lorentz transformations SO(3,1) contained in 
SO(3,2), and let $P$ denote the transformations of the (approximate) Poincar\'e 
group P(3,1).

Consider a reference object with state $|\Phi\rangle$ in a SO(3,2)-symmetric 
Hilbert space $H_S$.
Assume that in the neighbourhood $\cal{N}$ of the origin $\cal{O}$ this 
state is approximated by a momentum eigenstate.
When all Lorentz transformations $L$ are applied to this state,
a Hilbert space $H_L$, as a subspace of $H_S$ is obtained. 
This Hilbert space is associated with the point $\cal{O}$.

If a (finite) transformation $s \in S, s \not\in L$ is applied to a state 
of $H_L$, a new state is generated, which is not in $H_L$. 
Therefore, by applying transformations of the \begin{em}coset\end{em} $Ls$, 
a non-equivalent Hilbert space $H^s_L$ is obtained.
This Hilbert space is associated with the point $s\,\cal{O}$.
There is a one-to-one relation between cosets $Ls$ and Hilbert spaces
$H^s_L$.
The set of all cosets $Ls$ generates the total Hilbert space $H_S$.

The set of cosets forms a \begin{em}homogeneous space\end{em} $S/L$,
where $S$ acts transitive on this space and $L$ is the isotropy group
of the origin $\cal{O}$;
the projection $\pi:\,S \rightarrow S/L$ makes $S$ a principle bundle
on $S/L$ with fiber $L$. 

Adding up all non-equivalent $H^s_L$ means an integration over the
homogeneous space $S/L$.
The integral delivers a decomposition of $H_S$ into a sum of 
non-equivalent $H^s_L$
\begin{equation}
H_S = \int d\Omega\;H^s_L = \int ds\,\frac{d\Omega}{ds}\;H^s_L\;, \label{6-1}
\end{equation}
where $d\Omega$ is the infinitesimal volume element in $S/L$.
The Jacobian $d\Omega/ds$ is a measure of the number of 
non-equivalent Hilbert spaces $H^s_L$ obtained by an infinitesimal 
transformation $ds$.
With a properly chosen parameterisation, such that $\int \! ds = 1$, the 
Jacobian becomes identical to the volume $V(S/L)$ of $S/L$.
The volume of $S/L$ is determined in \cite{ggm1} as
\begin{equation}
V(S/L) = \frac{16 \pi}{3}\;.                              \label{6-7}
\end{equation}

Volumes of homogeneous spaces were calculated by L. K. Hua \cite{hl,lkh}.
They have been used with some success in semi-empirical mass formulae 
for more than three decades \cite{aw,fds}.

In the case of two or three reference objects, the Hilbert space $H_S$ is
obtained from product representations of individual Hilbert spaces
$H^{(1)}_S, H^{(2)}_S$, and eventually $H^{(3)}_S$. 
With the decomposition (\ref{6-1}) of each Hilbert space,
the integrals contain products of volume factors (Jacobians) $V(S/L)$.
The following factors correspond to one, two and three 
reference objects, respectively,
\begin{equation}
\left(\frac{16 \pi}{3}\right) \mbox{,} \;\;
\left(\frac{16 \pi}{3}\right)^2  \;\; \mbox{ and } \;\;
\left(\frac{16 \pi}{3}\right)^3.                         \label{6-9}
\end{equation}

Restricting the resulting Hilbert spaces to such Hilbert spaces that are 
associated with the point $\cal{O}$, means dividing the factors in 
(\ref{6-9}) by $V(S/L)$.
This results in
\begin{equation}
1  \mbox{ , } \;\;
\left(\frac{16 \pi}{3}\right)  \;\; \mbox{ and } \;\;
\left(\frac{16 \pi}{3}\right)^2  .                      \label{6-10}
\end{equation}

Next remember that spinors are attached to states of representations of 
P(3,1) in the neighbourhood $\cal{N}$ of $\cal{O}$, rather than of SO(3,1)
at $\cal{O}$.
Representations of P(3,1) in $\cal{N}$ are obtained from representations 
of SO(3,1) at $\cal{O}$ by adding infinitesimal transformations $t \in S$.
By applying all $t$ to $L$, cosets $Lt$ are obtained. 
They form a homogeneous space $P/L$ with a volume \cite{ggm} of
\begin{equation}
V(P/L)\;=\;V(U(1))\;=\;4\pi \;.                          \label{6-11}
\end{equation}
This corresponds to an additional integration over the homogeneous space 
$P/L$ with the Jacobian $V(P/L)$.

The Hilbert space $H_L$ (associated with the point $\cal{O}$) requires a 
special treatment. 
It is spanned by momentum eigenstates. 
In $\cal{N}$, an infinitesimal $t$ is equivalent to a translation.
Since momentum states are eigenstates of the generators of translations,
the effect of $t$ is a mapping of a momentum state onto the same state with 
an additional phase factor.
Since phase factors applied to basic states do not change the Hilbert space,
$H_L$ is not changed by translations and the factor $V(P/L)$ does not apply. 

This argument does not apply to other $H^s_L$ for the following reason.
$t$ does not commute with a finite $s$, except when $s$ is in the 
direction of $t$.
Therefore, the action of $t$ can in general not be described by   
phase factors applied to the base states of $H^s_L$. 
The special case, when $s$ is in the direction of $t$, delivers a contribution
to the integral of measure zero, which can be neglected.

Multiplying the terms in (\ref{6-10}) by the appropriate factors
results in the volume factors
\begin{equation}
1  \mbox{ , } \;\;
4\pi \left(\frac{16 \pi}{3}\right)  \;\; \mbox{ and } \;\;
4\pi \left(\frac{16 \pi}{3}\right)^2  .                 \label{6-14}
\end{equation}
This means, the involvement of two and three reference objects gives 
rise to a direct sum of non-equivalent Hilbert spaces $H_P$, determined 
by an integral over a parameter space and a Jacobian given by 
$(\ref{6-14})$.
We use here and in the following the term ``direct sum", although
it is in fact a ``direct integral".

The direct sum has formally the same structure as the Hilbert space of a 
multi-particle system, where the number of particles is determined by the 
multiplicity $(\ref{6-14})$.
Since the spinor mass has the same value $m_e$ in each $H_P$, the resulting 
effective spinor mass is $m_e$ multiplied by one of the volume factors 
(\ref{6-14}). 
This reproduces G-M's mass relations (\ref{6-8}).

There is a certain similarity to the Higgs mass generating mechanism:
Masses are determined by the coupling of spinors to the background spin 
network, which in some sense can be regarded as a background field.
However, this background field describes quasi-classical objects, which 
means that it is a classical field. 
Hence there are no ``Higgs particles".

The fact that the mass relations agree with experimental data, suggests 
an identification of the three configurations with massive leptons.

\section{Massless spinor states}

Assume that a spinor is coupled to a reference in such a way that for the 
resulting spinor momentum the relation $|p_0|=|$\boldmath$p$\unboldmath$|$ 
holds.
This relation defines a massless spinor state.
There are principally two configurations, a massless ``particle" with 
$p_0 > 0$ and a massless ``anti-particle" with $p_0 < 0$.
As long as the flat-space approximation is valid, all states can be generated
from a single particle or anti-particle state by a transformation of SO(3,1).
Remember that SO(3,1) does not include reflections.

A covariant description of these states is given by the Weyl equation,
which is known not to be invariant under space reflections,
\boldmath
\begin{equation}
\mbox{\unboldmath$(\partial_0\,+\,$} \sigma 
\mbox{\unboldmath$\cdot$} \partial \mbox{\unboldmath$ )\;\psi(x)=0$} 
\;\mbox{\unboldmath .} \label{7-1}
\end{equation}
\unboldmath
For positive $p_0$ the helicity 
\boldmath$\sigma$\unboldmath$\cdot$\boldmath$p$\unboldmath$\,/\,|$
\boldmath$\!\!p$\unboldmath$|$
is $-1$.
For negative $p_0$ (anti-particle) the helicity is $+1$.
This is consistent with the experimental fact that massless leptons exist 
only with these helicities.
Within our approach, the chirality of massless spinor states is a simple 
consequence of the fact that SO(3,1) does not include reflections. 

The spinor part of the solutions of the Weyl equation can also be written
as Dirac spinors (refer to standard text books)
\begin{eqnarray}
|u\rangle &=& \sqrt{E} \left(\begin{array}{*{1}{c}}
 0 \\ 1 \\ 0 \\ -1 \end{array}\right)  
 \; \mbox{ , } \;
|v\rangle  = \sqrt{E} \left(\begin{array}{*{1}{c}}
 0 \\ -1 \\ 0 \\ 1 \end{array}\right)     \;\;     .        \label{7-2}
\end{eqnarray}
In this form, massless leptons appear as a superposition of equally
weighted particle and anti-particle states of massive leptons.
Therefore, massless leptons do not really represent a new phenomenon.
Hence, the considerations of the preceding section, regarding the 
linking to one, two, or three reference objects, apply also to the solutions 
of the Weyl equation. 
Each massive lepton is then accompanied by a massless, neutrino-like 
configuration.
Consequently, we can identify \begin{em}three families of leptons\end{em}, 
caused by three types of referencing to macro-objects.

The question of how possibly non-zero neutrino masses can be obtained from
corrections based on the exact SO(3,2) symmetry is beyond the scope 
of this paper.

\section{Electromagnetic interaction}

Consider a spinor state in an SO(3,2) symmetric Hilbert space.
SO(3,2) transformations are then generated from infinitesimal generators 
that consist of an
``orbital" and a spin part
\begin{equation}
l_{\mu\nu} + s_{\mu\nu}  \;\;                          \label{8-1}
\end{equation}
and
\begin{equation}
l_{\mu4} + s_\mu   \; .                            \label{8-2}
\end{equation}
The operators (\ref{8-1}) generate the transformations of the
homogeneous Lorentz group.
In the neighbourhood of the point of contraction $\mathcal{O}$,
the operators (\ref{8-2}) are approximated by generators of ``translations"
\begin{equation}
p_{\mu} + s_\mu   \; .                             \label{8-3}
\end{equation}
Since there is no other spin matrix, which near $\mathcal{O}$ can be used 
as an approximation to $s_\mu$, the original matrix is used.
The momentum operator $p_\mu$ in (\ref{8-3}) stands for the effect of the 
spinor on the reference object.
The term $s_\mu$ in (\ref{8-3}) has to be considered as an inheritance from 
the global SO(3,2) symmetry.

In studying the contraction limit, the older literature, e.g.\cite{fg,ff}, 
concentrated on the case $|p| \gg |s|$.
Therefore, the contributions of $s_\mu$ were considered negligible compared
to those of $p_{\mu}$.
Here the occurrence of $s_\mu$ in (\ref{8-3}) demands for a more 
differentiated examination.
Although $p_\mu$ is obtained from the large momentum $P_\mu$ of the 
macroscopic system, the contributions of $p_\mu$ can not be considered large 
compared to those of $s_\mu$. 
In fact, since $p_\mu$ stands for the effect of $s_\mu$ on the macrosystem, 
\begin{em}both operators have the same magnitude\end{em}.
Therefore, whenever a spinor operator occurs together with the corresponding
momentum operator, it is advisable to have a careful look at the involved 
magnitudes.
 
In the neighbourhood of the point of contraction $\mathcal{O}$,
operators (\ref{8-3}) are used for infinitesimal translations of magnitude 
$\epsilon$ only. 
From 
\begin{equation}
[\,\epsilon s_\mu, \epsilon s_\nu\,] = -i \epsilon^2 \sigma_{\mu\nu}
                                                                \label{8-4}
\end{equation}
with $\epsilon \to 0$, follows, that in this context not only 
$p_\mu$ but also $s_\mu$ can be treated like commuting operators.
Also in the first terms of an expansion of an infinitesimal transformation 
\begin{equation}
I + \epsilon^\mu s_\mu + \dots                                   \label{8-5}
\end{equation}
all higher terms may be neglected compared to the unit matrix $I$.
Therefore, it is correct to say that translations applied to the solutions 
of the Dirac equation are generated by $p_\mu$ only.
But ignoring $s_\mu$ in the general expression (\ref{8-3}), without any
justification, would be definitely a serious fault. 

Now consider an \begin{em}isolated\end{em} SO(3,2) based 
multi-particle system in flat-space approximation.
In a Poincar\'e invariant theory the word ``isolated", applied to a 
non-interacting multi-particle system, usually means that the 
modulus of the total 4-momentum is a constant
\begin{equation}
(\sum_i p_i^\mu)(\sum_j p_{j\mu}) = \mbox{const} \; .          \label{8-6}
\end{equation}
If the system interacts, we would in addition require that the
interactions are confined to members of the system.

In generalizing these remarks to the flat-space approximation of our SO(3,2) 
invariant theory, it seems reasonable to study configurations, where the 
modulus of the sum of individual generators (\ref{8-3}) is a constant 
\begin{equation}
[\sum_i (p_i^\mu + s_i^\mu)][\sum_j (p_{j\mu} + s_{j\mu})]=\mbox{const}\;. 
                                                                 \label{8-7}
\end{equation}
Based on the translation invariance of the flat space approximation, in 
addition to (\ref{8-7}), relation (\ref{8-6}) must still be valid.
This allows to write down another constant expression by eliminating   
contributions of (\ref{8-6}) to (\ref{8-7})
\begin{equation}
\sum_{ij} p_i^\mu s_{j\mu} + \sum_{ij} s_i^\mu s_{j\mu} =\mbox{const} \;. 
                                                                 \label{8-8}
\end{equation}
The terms $i=j$ in the first sum of (\ref{8-8}) are elements of Dirac's 
equations for individual spinors.
The terms $i\not=j$ define correlations between momentum and spin of 
different spinors.
The terms $i\not=j$ in the second sum define correlations between different 
spinors. 
For $i=j$ these terms simply deliver a constant unit matrix. 
Equation (\ref{8-8}) can be understood as a generalization of Dirac's 
equation to a multi-particle system.

It will become clear later that the constancy of the left hand side of 
(\ref{8-8}) is equivalent to the clause that interactions within the system 
are confined to members of the system.

To study the implications of (\ref{8-8}), a perturbation approach seems
appropriate.
This is suggested by those terms of (\ref{8-8}) that belong to 
Dirac's equation of individual spinors, similar to a non-interacting 
Poincar\'{e} invariant theory.
When we start from a system of ``free" Dirac particles as a zeroth 
approximation, we have to treat the other elements of (\ref{8-8}) as 
perturbations.
It is fairly evident from the structure of (\ref{8-8}) that such an approach
will introduce correlations between the originally ``free" particles.
In fact, we will find that these correlations take on the structure of 
well-known interaction terms.  

Let us start with the first term of (\ref{8-8}), and postpone the treatment
of the second term to section 10.

To treat the multi-particle system we make use of standard Fock space 
methods. 
The `free' part of the system is easily converted into a Fock space 
formulation for solutions of the Dirac equation. 
We skip this step and refer to standard textbooks (see e.g. \cite{gs}).
The field operator of the Dirac field (taken from this reference) has
the form
\begin{equation}
\psi(x) = (2\pi)^{-3/2} \!\!\int\!\! d^3p 
\left( b_s({\mathbf{p}}) u_s ({\mathbf{p}}) e^{-ipx} 
\!+\! {d_s({\mathbf{p}})}^\dagger v_s({\mathbf{p}}) 
e^{ipx} \right).                                        \label{9-1}
\end{equation}
A similar expression defines the adjoint operator $\psi^\dagger(x)$. 
$b^\dagger_s({\mathbf{p}}), b_s({\mathbf{p}})$ are electron emission 
and absorption operators, $d^\dagger_s({\mathbf{p}}), d_s({\mathbf{p}})$ 
are the corresponding operators for positrons.
They satisfy the usual anticommutation relations of the Dirac field.

Now let us try to find a Fock space representation for the 
correlation terms
\begin{equation}
s_\mu\, p^\mu \;                                   \label{8-9}
\end{equation}
in (\ref{8-8}).
The $s_\mu$ part is easily converted into a Fock space operator
\begin{equation}
\int\! d^3x \; \bar{\psi}(x)\,\gamma_\mu\, \psi(x) \; ,     
\end{equation}
where $\bar{\psi} = \psi^\dagger \gamma_0$ is the Dirac adjoint field 
operator.
When $\gamma_\mu$ belongs to a first particle, then the momentum $p^\mu$ 
belongs to a second one.
Adding a similar Fock space representation for $p^\mu$, would leads to a 
perturbation term, that is highly non-linear in $\psi(x)$:
\begin{equation}
\int\! d^3x \; \bar{\psi}(x)\,\gamma_\mu\, \psi(x) \; 
\bar{\psi}(x)\, p^\mu\, \psi(x) \; .     \label{8-10}
\end{equation}
A well-established method to handle non-linear many-body problems is the 
introduction of a \begin{em}potential\end{em} that is generated by the second 
particle and acts on the first, and vice versa.
In our case, the second part of (\ref{8-10}), which is related to the momentum
of the second particle, offers itself to be replaced by a place-holder operator 
$A^\mu(x)$. 
This operator will be defined as an operator in an auxiliary Fock space.
The states of this Fock space will formally be treated like independent 
degrees-of-freedom. 
However, the perturbation algorithm will be set up in such a way that 
conservation of 4-momentum is guaranteed at each ``vertex".
This condition will ensure that, within the algorithm, $A^\mu(x)$ will 
behave in the same way as the original operator (\ref{8-10}) would do at 
each vertex, as far as the balance of 4-momentum is concerned.

So let us try to write the correlation operator in the following form
\begin{equation}
\int\! d^3x \; \bar{\psi}(x)\gamma_\mu \psi(x) \, A^\mu(x) \, , \label{9-2}
\end{equation}
where the Fock space operator $A^\mu(x)$ has the effect of a quantized vector 
potential.
This potential shall describe the action of the second particle within the 
correlation term.

We make an ansatz for $A^\mu(x)$ in a general form, which is 
well-known from quantum electrodynamics (see e.g. \cite{gs}), 
\begin{eqnarray}
A^j(x) = (\frac{1}{2\pi})^{-3/2} \!\!\int\!\! \frac{d^3k}{k^0 \sqrt{2}} 
\left( a^j({\mathbf{k}}) e^{-ikx} 
+ a^j({\mathbf{k}})^\dagger e^{ikx} \right), \nonumber \\
j=1,2,3,                                                     \label{9-8}
\end{eqnarray}
and
\begin{equation}
A^0(x) = (\frac{1}{2\pi})^{-3/2} \!\!\int\!\! \frac{d^3k}{k^0 \sqrt{2}} 
\;i \left( a^0({\mathbf{k}}) e^{-ikx} 
+ a^0({\mathbf{k}})^\dagger e^{ikx} \right).                 \label{9-9}
\end{equation}
The operators $a^\mu({\mathbf{k}})$ and their counterparts 
$a^{\mu\dagger}({\mathbf{k}})$ act as absorption and emission operators for 
quanta with momentum $\mathbf{k}$.
As such they satisfy the commutation relations
\begin{equation}
[\,a^\mu({\mathbf{k}}), a^\nu ({\mathbf{k'}})^\dagger] 
= \delta^{\mu\nu}\delta({\mathbf{k - k'}}) \, .               \label{9-7}
\end{equation}
In (\ref{9-8}) and (\ref{9-9}) $k^0$ shall be determined by the requirement 
of energy-momentum conservation at the vertex, when these operators are 
evaluated within a matrix element.
In QED, where $A^\mu(x)$ is interpreted as the massless photon field, this
requirement leads to an `off-shell' behaviour of the photon field.
In our context, `off-shell' is not defined, since we have not assigned any 
mass to the place-holder field. 

The auxiliary Fock space can be understood as a bookkeeping mechanism for 
state transitions of the second particle.
The momentum operator $p^\mu$ in (\ref{8-11}) indicates that 
$a^\mu ({\mathbf{k}})^\dagger$ and $a^\mu({\mathbf{k}})$ not only create 
and annihilate states that carry momentum, but themselves contribute to
the balance of 4-momentum, when evaluated between states of the auxiliary 
Fock space.

Using the decomposition of the field operators (\ref{9-1}), (\ref{9-8})
and (\ref{9-9}) into their counterparts in momentum space, the contributions 
to (\ref{9-2}) take on the form
\begin{equation}
\dots\; {\bar{b}({\mathbf{p + k}}) \, \gamma_\mu \, b({\mathbf{p}})} \, 
a^\mu({\mathbf{k}}) \dots                             \label{9-4}
\end{equation}
(For a while, we omit the factors $u_s$, $v_s$ and the time dependencies.)
An analogous consideration is valid for positron and mixed terms.
Here $a^\mu({\mathbf{k}})$ has to simulate terms belonging to the second
particle of the form
\begin{equation}
\dots\; {\bar{b}({\mathbf{p' + k}}) \, p^\mu \, b({\mathbf{p'}})} \, \dots .
\label{8-11}                          
\end{equation}
Obviously we can integrate over $p'$ and obtain an expression for
$a^\mu({\mathbf{k}})$ in terms of pairs of annihilation and creation operators
of the second particle
\begin{equation}
a^\mu({\mathbf{k}}) = \int_V d^3({\mathbf{p'}})\, j_V(p')\;\dots\; 
{\bar{b}({\mathbf{p' + k}}) 
\dots b({\mathbf{p'}})} \, \dots  \; .                     \label{8-12}                          
\end{equation}
Here $V$ denotes the integration volume and $j_V$ is the Jacobian belonging
to this volume.
The integration volume will be evaluated in the next section.

Coming back to the expression (\ref{9-4}), we now add space-time 
dependencies to the emission and absorption operators  
\begin{equation}
\dots \; \bar{b}({\mathbf{p+k}})\, e^{i(p+k)x}\; \gamma_\mu \;
b({\mathbf{p}})\, e^{-ipx}\,\, 
a^\mu({\mathbf{k}})\, e^{-ikx} \dots \;.                       \label{9-10}
\end{equation}

After inserting the spin functions $u_s(\mathbf{p})$ and $v_s(\mathbf{p})$,
these terms and the corresponding positron and mixed terms add up to a 
Fock operator in the form
\begin{equation}
\int d^3x\, : \bar{\psi}(x)\gamma_\mu \psi(x) : A^\mu(x) \;, \label{9-11}
\end{equation}
where the dots $::$ stand for normal ordering of emission and absorption 
operators.
This operator has the form of the interaction term of quantum 
electrodynamics (QED).
Therefore, we can conclude that (massive) spin-1/2 particles, based on 
SO(3,2) symmetry, carry a charge and interact according to the rules of QED. 

The coupling constant is hidden in the Jacobian $j_V$, which, obviously,
relates transitions ${\mathbf{p'}} \rightarrow {\mathbf{p'+k}}$ to the 
generation or absorption of a ``photon" with momentum ${\mathbf{k}}$.
Within a perturbation calculation another Jacobian belongs to the 
integration over ${\mathbf{k}}$.
Our task is therefore, to adjust the Euclidean volume element $dk\,dp$
to the geometry of the integration volume by a combined Jacobian.
This volume has, as we shall see below, an essentially spherical structure.
The Jacobian can be understood as the density of states that contribute to 
an Euclidean volume element. 
Therefore the evaluation of the Jacobian in the following section will 
essentially consist on an evaluation of this density.

\section{Calculation of the coupling constant}

The determination of the Jacobian is, by no means, a trivial task. 
Nevertheless, it was (unwittingly) attacked 40 years ago by 
A. Wyler \cite{aw}, 
who discovered that the fine-structure constant $\alpha$ can be expressed
by volumes of certain symmetric spaces. 
Unfortunately, Wyler was not able to put his observation into a convincing 
physical context. 
Even worse, his mathematical reasoning did not withstand a closer 
inspection.
Therefore, his work was criticized as fruitless numerology \cite{br}.  
We will see in the following that Wyler's approach has a very down-to-earth 
mathematical basis.
But neither Wyler nor his critics were aware of this. 

As a matter-of-fact Wyler broke down the problem of determining the Jacobian
into several steps, by skilfully making use of some symmetric structures.
These structures have well-known volume and surface elements, which Wyler
then put together to obtain his formula.

To retrieve these symmetric structures, we take a close look at the parameters
that describe particle states in momentum representation.
The state of a single particle depends on three independent parameters 
$p_1,p_2,p_3$ with
\begin{equation}
p_0^2 -p_1^2 - p_2^2 - p_3^2 = m^2 . \label{10-1}
\end{equation} 
They span a 3-dimensional parameter space. 
Two particle momenta then span a 6-dimensional parameter space.
When the momenta of both particles are given, we can use (\ref{10-1}) to 
calculate the effective mass of the two-particle state.
The value of the mass characterizes an irreducible representation of
P(3,1). 
When we hold this mass fixed, the number of independent parameters is 
reduced to five.
Let us denote these parameters by $z = (z_1,z_2,z_3,z_4,z_5)$, $z \in R^5$. 

Next consider the Dirichlet problem for Laplace's equation in $R^5$.
The Dirichlet problem consists in finding a solution $\phi(z)$ on some
domain $D$, such that $\phi$ on the boundary of $D$ is equal to some given
function.
Following Wyler, we extend the parameter space into the complex space $C^5$ 
and make use of the Poisson kernel $P(z,\xi)$, which is 
defined on a polydisk $D^5$ in $C^5$.
We choose the radius of $D^5$ large enough, so that all momenta, that for a 
given two-particle state may contribute to (\ref{8-12}), are enclosed.
Then we scale the momentum parameter, so that the largest momentum has the 
value of 1.
Therefore, without loss of generality, we can set the radius of the polydisk
to 1, as Wyler did, and write 
\begin{equation}
D^5 = \{z \in C^5; 1 + |zz'|^2-2\bar{z}z' > 0, |zz'| < 1 \}\;. \label{10-8}
\end{equation}
The Poisson kernel has the well-known reproducing property that 
analytic functions $\phi(z)$ in $D^5$ are determined by their values 
on the boundary $Q^5$ of $D^5$
\begin{equation}
\phi(z) = \int_{Q^5} P(z,\xi) \phi(\xi) d\xi \; , \label{10-9}
\end{equation}
where
\begin{equation}
Q^5 = \{\xi = xe^{i\theta}; x \in R^5, xx'=1 \}\;. \label{10-10}
\end{equation}
This integral will be used as a vehicle to study the density of states 
that contribute to the Euclidean volume element $d\xi$.

The Poisson kernel is given in \cite{lkh} by
\begin{equation}
P(z,\xi) = \frac{1}{V(Q^5)} \frac{(1+|zz'|^2 - 2\bar{z}z')^\frac{5}{2}}
{|(z-\xi)(z-\xi)'|^5} \; .                          \label{10-11}
\end{equation}
For $z=0$ the Poisson kernel takes on a constant value 
\begin{equation}
P(0,\xi) = \frac{1}{V(Q^5)} \; ,                      \label{10-12}
\end{equation}
and, therefore, from (\ref{10-9})
\begin{equation}
\phi(0) = \int_{Q^5} P(0,\xi) \, \phi(\xi) d\xi 
=  \int_{Q^5} \phi(\xi) \frac{d\xi}{V(Q^5)} \;.      \label{10-13}
\end{equation}
Each volume element on $Q^5$ then contributes with a factor of $V(Q^5)^{-1}$
to the integral.

The five dimensions of $Q^5$ may appear to indicate that there is a 
symmetry with respect to SO(5).
This is not the case.
A rotation from the momentum subspace of the first particle to 
the second is not a valid symmetry operation and hence cannot
contribute to the states in a volume element.
Excluding \begin{em}one\end{em} axis of rotation from the symmetry operations
of $Q^5$ means a reduction of the symmetry volume from five to four 
dimensions.
We therefore have to correct the integration volume by a factor of 
$1/V(S^4)$, where $S^4=SO(5)/SO(4)$ is the unit sphere in 4 dimensions.
We account for this correction by multiplying the volume element of the 
Poisson integral by this factor. 
This delivers a new value of $1/(V(Q^5)\,V(S^4))$ for the density of  
states on $Q^5$.

Up to now the integral covers states, whose parameters are located on 
the boundary $Q^5$.
To include states, whose parameters are located in the interior of $D^5$,
we have to extend the integral by an integration in a radial direction, e.g. 
in the direction of $z_1$.
It is sufficient to integrate over \begin{em}one\end{em} radial direction, 
because the other directions are already covered by the integration over 
$Q^5$.
(Remember, that the integration over $Q^5$ has indeed been performed, as if 
there were a SO(5) symmetry. 
The error made in this connection has been compensated
for only after the integration.)

In preparation for this integration, let us consider the integral
\begin{equation}
\int_0^1 dr \int_{Q^5} \frac{d\xi(r)}{V(Q^5)}  \;, \label{10-14}
\end{equation}
where $r$ is the radius of $Q^5$.
This integral delivers the volume $V(D^5)$ of $D^5$.
Instead of integrating over $r$, we can multiply the volume element of
$Q^5$ by $V(D^5)$ and evaluate the following integral
\begin{equation}
\int_{Q^5} \frac{V(D^5)}{V(Q^5)}\,d\xi  \;,        \label{10-15}
\end{equation}
which delivers the same result, since
\begin{equation}
\int_{Q^5} \frac{1}{V(Q^5)}\,d\xi = 1 \;.          \label{10-16}
\end{equation}

Since the volume element $d\xi(r)$ is 4-dimensional, the integration 
in (\ref{10-15}) is performed in four orthogonal directions.
Therefore, an integration over \begin{em}only one\end{em} direction 
can be replaced by a multiplication of the volume element by 
$V(D^5)^\frac{1}{4}$.
Hence, the integration over a radial direction in $D^5$ contributes an 
additional factor $V(D^5)^\frac{1}{4}$ to the volume element $d\xi$.

When we perform the integration over the phase factor in (\ref{10-10}),
which can be done for ``reasonable" $\phi(z)$ (cf. \cite{lkh}),
we obtain another factor of $2\pi$.

Remembering that $2 \times 2$ spin components contribute to a two-particle 
state, we add a factor of 4.

Collecting all factors results in an effective volume factor of
\begin{equation}
8 \pi \,V(D^5)^{\frac{1}{4}} \, / \, (V(S^4) \, V(Q^5)).    	\label{10-28}
\end{equation}
This is Wyler's formula.

The volumes $V(D^5)$ and $V(Q^5)$ have been calculated by 
L. K. Hua \cite{lkh}. 
$V(S^4)$ is the volume of the unit sphere $S^4$ in 4 dimensions. With
\begin{equation}
V(Q^5) = \frac{8 \pi^3}{3},        			\label{10-29}
\end{equation}
\begin{equation}
V(D^5) = \frac{\pi^5}{2^4\, 5!},   			\label{10-30}
\end{equation}
\begin{equation}
V(S^4) = \frac{8 \pi^2}{3}         			\label{10-31}
\end{equation}
we obtain 
\begin{equation}
\frac{9}{8 \pi^4} \left(\frac{\pi^5}{2^4 \, 5!}\right)^{1/4}   	
 = \; \frac{9}{16 \pi^3} \left(\frac{\pi}{120}\right)^{1/4} 
= \; 1/137.03608245.   			          	\label{10-32}
\end{equation}
The best experimental value for $\alpha$ currently is 1/137.035 999 070(98).

Wyler's description of the factor $V(D^5)^\frac{1}{4}$ is not very clear, but 
his formulations indicate that he regarded this additional volume factor as 
the Jacobian of the boost operation $z \rightarrow g(z)$.
From the explicit form of the Poisson kernel (\ref{10-11}) we can easily read 
off that this interpretation cannot be correct.
Nevertheless, Wyler can take the credit for having found a precious key to a
basic aspect of particle physics.
It was his personal fate that he did not find the matching lock.

The correct value for the fine-structure constant delivers strong evidence 
that the interaction, found in the last section, is in fact the 
electromagnetic interaction.

\section{Strong interaction}

After having successfully reproduced the electromagnetic interaction,
we should feel encouraged to look for other interactions that may result from
SO(3,2) corrections.
This and the next two sections will briefly sketch, how these interactions are 
obtained.
We start with an interaction term that is very similar to the electromagnetic 
term, but is obtained from a different configuration of spinors.

So far we have treated the linkage of a \begin{it}single\end{it} spinor to up
to \begin{it}three\end{it} reference objects.
Now we will try the opposite situation, where \begin{it}three\end{it} 
spinors are linked to a \begin{it}single\end{it} quasi-classical reference 
object.

The action of the cluster of three spinors on the reference is again a change
of its 4-momentum, which will be interpreted as the 4-momentum of the cluster.
Let us assume that the cluster of spinors represents a stable configuration 
with respect to transformations of the Poincar\'e group. 
Then there are three independent momentum components of the cluster that have 
to be distributed among three spinors.
To solve this task, we give the operators of the Dirac field an additional
index $\psi_k(x)$, where $k=1,2,3$ references one of the three momentum 
components.
With the help of Gell-Mann's matrices $\lambda_a, a=1,\dots,8$, we can define
eight basic configurations 
\begin{equation}
\sum_{k=1}^3 \lambda_{a,ik} \, \psi_k(x) \, , \;\;a=1,\dots,8 \; .\label{IX-1}
\end{equation}
From the octet of Gell-Mann's matrices any other $3\times 3$ matrix with trace
0 can be obtained by linear combination.
In this sense the configurations (\ref{IX-1}) form a complete basis.
When we insert this ansatz into the interaction term (\ref{9-2}), and perform
the same steps as in section 7, we obtain the interaction term of
quantum chromodynamics
\begin{equation}
\int\! d^3x \; \bar{\psi}(x)\gamma_\mu \lambda_a \psi(x) \, G_a^\mu(x) \; . 
                                                           \label{IX-2}
\end{equation}
Here the Gluon field $G_a^\mu(x)$ takes the place of the electromagnetic 
four-potential $A^\mu(x)$.

A quark appears as a kind of uncompleted particle in the sense that
it does not possess a complete set of momentum components relative to a 
reference.
A single quark may contribute just one component, which may be either 
$p_1$, $p_2$ or $p_3$.
Therefore, it is not possible to give a single quark a position relative to 
a reference.
When we associate the colours red, green or blue to each of the momentum 
components, then only a ``white" combination provides a complete momentum 
vector. 
This then allows to treat the compound object as a free particle. 
Since a quark can carry any of the three momentum components, there is
a SU(3) symmetry with respect to exchange of the momentum components.

\section{Weak interaction}

When we write the second term of (\ref{8-8}) as a Fock space operator 
and specialize this operator to the transition 
$\mu \rightarrow e + \nu_\mu + \bar{\nu_e}$, we obtain the basic structure 
of weak interaction
\begin{equation}
\int\! d^3x \; \bar{\psi}_{\nu_\mu}(x)\,\gamma_{\mu L}\,\psi_{\mu}(x) \;\, 
\bar{\psi_e}(x)\,\gamma^\mu_L\,\psi_{\nu_e}(x)  \; .              \label{X-1}
\end{equation}
The matrices $\gamma_{\mu L}$ contain the usual projection on left-handed 
components
\begin{equation}
\gamma_{\mu L} := \gamma_\mu \frac{1 - \gamma_5}{2} \; ,  \label{X-2}
\end{equation}
since, as we have seen in section 6, there are only left-handed neutrinos.

\section{Gravitation}

In a SO(3,2) symmetric system, curved space-time obviously is an inherent 
property.
Flat space-time is obtained only as an approximation.
Up to now we have tacitly assumed that this approximation is valid.
In the following we will try to find configurations, where this assumption  
is not valid any more.

Consider the SO(3,2) operator (\ref{1-7}) from section 2 
\begin{equation}
j_{\mu4} = i(x_\mu\, \partial_4 - x_4\, \partial_\mu),\;\;\; \mu = 0,1,2,3 \;. 
\label{XI-1}
\end{equation}
In tangential space-time at the point $O=(0,0,0,0,R)$ the translation 
operator is proportional to the second part $-i x_4\, \partial_\mu$ of the 
SO(3,2) operator.
The contribution of the first term will, in general, be very small, so that it 
can be neglected.
But, when we examine SO(3,2) correction to the flat-space approximation, 
we must take care of the first part $i x_\mu\, \partial_4$ in a suitable way.

When we add the first part as a ``small" correction to the translation 
operator, in the same way as we have added correction terms before, we 
encounter product terms of the form
\begin{equation}
i\,x^\mu\,\partial_4 \; (-i\,x'_4\,\partial'_\mu 
- i\,x''_4\, \partial''_\mu - \dots) \;,                    \label{XI-2}
\end{equation}     
which result from an evaluation of (\ref{8-7}) with $p_\mu$ replaced by
$j_{\mu 4}$.
Here the non-primed term on the left refers to a test particle under 
examination and the primed terms in brackets shall refer to a large collection
of other particles in the neighbourhood of the test particle.
We assume that the primed particles are concentrated near $O$, so that 
$x'_4, x''_4, \dots$ are nearly equal to $R$.
When, furthermore, the wave functions of the primed momenta are in phase, 
the expectation values of the momentum of the primed particles can add up to a 
large value.
The largeness of this value may then eventually over-compensate the smallness 
of the term $x_\mu\,\partial_4$. 
Thus, in a perturbation calculation the contribution of (\ref{XI-2}) may
become dominant. 

To understand the effect of (\ref{XI-2}) in a perturbation expansion, we 
replace $x_4$ in (\ref{XI-1}) by the de Sitter radius $R$, divide by $R$ 
and obtain
\begin{equation}
i\,\frac{x_\mu}{R} \, \partial_4 \,  - i\,\partial_\mu \;.     \label{XI-3}
\end{equation}
This operator stands for a de Sitter displacement operation near $O$. 
For $R \to \infty$ this operation becomes the translation operator.
For finite $R$ it describes the corresponding operation in curved space-time
with a radius of curvature $R$.

Now compare (\ref{XI-3}) with the term 
\begin{equation}
i\,x^\mu \,\partial_4 \; (-i\,\partial'_\mu+\dots ) 
- i\gamma^\mu \partial_\mu \;,                                 \label{XI-4}
\end{equation}     
obtained by inserting (\ref{XI-2}) into a perturbation expansion for a test 
particle near the point $O$.
It is obvious that (\ref{XI-4}) has the characteristics of the de Sitter 
displacement operator (\ref{XI-3}) in curved space-time.
But now the curvature $1/R$ is replaced by the 4-momentum of primed particles.
Therefore, the curvature of space-time, experienced by the test particle,
becomes proportional to the sum of 4-momentum of primed particles.
Taking into account a distribution of the primed particles over a finite area, 
we can (in a classical approximation) replace their 4-momentum by an 
energy-momentum tensor.

It may be of interest that, even in the absence of any other matter, 
space-time of the test particle is curved as indicated by (\ref{XI-3}), 
due to the basic SO(3,2) symmetry.

When Albert Einstein set up the field equations of general relativity, he used
the fact that the relation between curvature, expressed by the metric tensor 
$g_{\mu \nu}$, and the energy-momentum tensor is uniquely determined by 
the following conditions \cite{ae}:

1. There are no higher than second order differential quotients of 
$g_{\mu \nu}$.

2. The relation is linear in the second order differential quotient.

3. The divergence of the curvature expression is zero.

Although further work will be needed to definitely show that these conditions 
are satisfied for the quantum mechanical interaction term (\ref{XI-2}), we 
can state the following:
The curvature, experienced by the test particle, is proportional to the
energy-momentum tensor of the particles in its neighbourhood.
This proportionality has the following consequences:
Since the curvature tensor is linear in the second order differential 
quotient of $g_{\mu \nu}$, condition 2 is obviously satisfied.
The curvature tensor does not contain differential quotients of greater then 
second order, therefore also condition 1 is satisfied.
Condition 3 expresses energy-momentum conservation. 
Since energy-momentum of the primed particles is conserved, at least in a 
neighbourhood of $O$, condition 3 also holds.

The aforesaid strongly suggests that the term (\ref{XI-4}), within a
perturbation calculation, defines a theory of quantum gravity that, 
in the classical (low-energy) limit, reproduces the field equations of 
general relativity.

The reader will have noticed that the term (\ref{XI-4}) does not contain 
any ``gravitational field".
It only contains well-defined differential operators acting on wave functions 
of the involved particles.
Therefore, the notorious problems of ``quantizing" the ``gravitational field"
are avoided.

\section{Spin networks versus binary elements}

The approach has started from a spin network, although its motivation
came from information at its most elementary level, represented by a set
of binary elements.
So the author owes the reader a justification of this proceeding.

Consider a set of binary elements, representing information.
The information may be complete, but, in general, it is not.
Therefore, we have to provide the possibility of representing incomplete 
information.
A simple method to do this is to form linear combinations with complex 
coefficients, which interpolate between the states of the binary element.
This leads to a vector space, which can easily be extended to a Hilbert space.
By this action a binary element is extended to a spinor.
The spinor form can be considered as the quantum mechanical representation 
of a binary element.

As before we can use these spinors to construct a quasi-classical 
reference.
After having done this, we are in a position to look for a covariant 
description of a single binary element \begin{em}relative\end{em} to the 
reference. 
Since a rotation of the reference, together with its coordinate system, 
induces a similar rotation of the spinor, the task of finding a covariant
description is settled by the spinor form of the binary element. 
Then spin networks are ``relational equivalent" to sets of binary
elements. 

This consideration shows that Bohr's statement not only can be applied to 
the most elementary level of information, but actually can be regarded as 
the root of quantum mechanics at this basic level.

\section{Conclusion}

Bohr's statement on the relational nature of observation has guided us from
a most general binary structure, to concepts that in the end, allow us to
describe observations relative to experimental setups in the common language 
of particle physics. 

Bohr's statement has directed us to a description of 
\begin{em}microscopic phenomena relative to macroscopic setups\end{em} 
by interpolating Hilbert spaces. 
Therefore, quantum mechanics can be understood as a necessary and logical 
result of the application of Bohr's statement to the microscopic domain.

Another consequence of this approach is the emergence of Minkowskian 
space-time.
Space-time is obtained as the parameter space of local symmetry
transformations of a quasi-classical reference object.
It is then employed to describe binary elements relative to 
quasi-classical reference objects in a covariant way.
Since SO(3,2) contains the translation group only as a local approximation, 
we observe corrections to translation invariance, that manifest themselves
as interactions between certain configurations of spinors.

Note that we have obtained space-time as a property of 
more or less \begin{em}macroscopic\end{em} objects. 
This is polar opposite to attempts to derive physical space-time
from discrete structures at Planck scales.
Nevertheless, our approach is basically also a discrete one,
but discrete at microscopic scales rather than at Planck scales.

``Internal" symmetries, considered as fundamental in the standard model, like 
SU(3), SU(2) and U(1) have been found to emerge as symmetries of special 
spinor configurations.
 
In contrast to the standard model there are no adjustable constants.
This means that, when spin-networks are taken as a basis for the
standard model, its constants must be calculable from some 
characteristic configurations of spinors.
That this is possible in principle, has been shown for the lepton mass 
spectrum and the fine-structure constant.

Our approach exhibits a general binary structure as a basis, from which 
fundamental aspects of particle physics can be derived.
Does this imply that reality has the structure of a binary system?
Is this structure some kind of a pre-physical aspect of reality, or is it 
rather a logical consequence of our continuous search for more and more 
elementary structures?
In the author's opinion the latter is true.
The binary structure reflects our way of doing science along the concept of
reductionism. 
Reductionism applied to information must, inevitably, lead to the most 
elementary, which means binary, elements of information.
Since binary information cannot be divided further, there cannot be any 
knowledge beyond what can be coded into the binary structure.
Binary structures, therefore, do not require or even allow for a deeper 
explanatory base.
The binary structure serves as a general means to represent knowledge at 
the most elementary level, but it is not part of the knowledge. 
It is merely an unspecific logical scheme, comparable to the binary system in
computer technology, which is used only to \begin{em}represent\end{em}
information.

After having divided knowledge into their most elementary parts, it 
should not be a surprise that we can recombine the parts and 
obtain space-time, particles and interaction.
What could come as a surprise is that we can do this without the help of any 
``law of nature". 
In fact, the only principle that we have applied is formulated in Bohr's 
dictum about the relativity of observation.
This means that quantum mechanics, space-time, particles and interactions are 
inevitable consequences, not of fundamental laws of nature, but of 
how we observe physical phenomena.
The transcription of our observation techniques into a mathematical form does 
not reveal any deep mystery, but only employs the rules of mathematical logic.

Bohr may have had something near to it in mind, when he said: 
``It is the task of science to reduce deep truths to trivialities".

\renewcommand{\baselinestretch}{1.1}


\begin{thebibliography}{99}

\bibitem{ws1} W.~Smilga, ``Elementary informational structures and their
relation to quantum mechanics and space-time", available as physics/0502040.

\bibitem{cr} C. Rovelli, ``Relational Quantum Mechanics", 
Int. Jour. Mod. Phys. 35, 1637-1678 (1996).

\bibitem{rl} R. Laughlin, ``A Different Universe: Reinventing Physics 
from the Bottom Down" (Basic Books Inc., 2005) 

\bibitem{bh} Cited in B.~Herbert,
\begin{em}Quantum Reality\end{em}, p. 45,
(Anchor Press, Garden City, N.Y., 1985).

\bibitem{lfcr} Laudisa, Federico and Carlo Rovelli,
``Relational Quantum Mechanics",in: Edward N. Zalta, Ed.,
\begin{em}The Stanford Encyclopedia of Philosophy\end{em},
(Spring 2008 Edition), \\
http://plato.stanford.edu/archives/spr2008/entries/qm-relational/. 

\bibitem{jaw} J. A. Wheeler,
``Information, physics, quantum: The search for links"
in: W. H. Zurek, Ed.,   
\begin{em}Complexity, Entropy and the Physics of Information\end{em}, 
Addison-Wesley, 1991

\bibitem{rp} R. Penrose, ``Angular momentum: an approach to 
combinatorial space-time", in: \begin{em}Quantum Theory and Beyond\end{em},  
ed. Ted Bastin (Cambridge University Press, Cambridge, 1971).

\bibitem{rpwr} R. Penrose, W. Rindler, 
\begin{em}Spinors and space-time\end{em}, p. 43,
(Cambridge University Press, Cambridge, 1988)

\bibitem{cfw1} C. F. von Weizs\"acker, ``Binary Alternatives and Space-Time
Structure", in: Proceedings, Quantum Theory and the Structure of Time and
Space, Vol.2 (In Memoriam Werner Heisenberg) (Munich, 1977).

\bibitem{cfw2} C. F. von Weizs\"acker, ``Urs, Particles, Fields", 
in: Proceedings, Quantum Theory and the Structure of Time and 
Space, Vol.5 (Munich, 1983).

\bibitem{crls1} C. Rovelli and L. Smolin, ``Spin Networks and Quantum 
Gravity", Phys. Rev. D 52, 5743-5759 (1995).

\bibitem{crls2} C. Rovelli and L. Smolin, ``Discreteness of area and 
volume in quantum gravity", Phys. Rev. B 442, 593-622 (1995).

\bibitem{mj} As quoted in M.~Jammer, 
\begin{em}The Philosophy of Quantum Mechanics\end{em} 
(Wiley, 1974)

\bibitem{iw} E. In\"on\"u and E. P. Wigner, 
Proc. Nat. Acad. Sci. USA \bfseries 39\mdseries, 510 (1953).

\bibitem{fg} F. G\"ursey in: \begin{em}Group Theoretical Concepts and
Methods in Elementary Particle Physics\end{em}, ed by G\"ursey  
(Gordon and Breach, 1964). 

\bibitem{ggm1} G. Gonz\'alez-Mart\'in, ``p/e Geometric Mass Ratio",
Reporte SB/F/278-99, Univ. Sim\'on Bol\'ivar (1999);
available as physics/0009066.

\bibitem{ggm} G. Gonz\'alez-Mart\'in, ``Lepton and Meson Masses",
Reporte SB/F/304.4-02, Univ. Sim\'on Bol\'ivar (2003);
available as physics/0405094.

\bibitem{hl} L. K. Hua and K. H. Look (= Lu, Qi-keng),
Scientia Sinica \bfseries8\mdseries, 1031-1094 (1959).

\bibitem{lkh} L. K. Hua, \begin{em}Harmonic Analysis of Functions of
Several Complex Variables in the Classical Domains\end{em},
(American Mathematical Society, Providence, 1963).

\bibitem{aw} A. Wyler,
C. R. Acad. Sc. Paris \bfseries271A\mdseries, 186-188 (1971).

\bibitem{fds} F. D. Smith, Jr.,
Int. J. Theor. Phys. \bfseries24\mdseries, 155 (1985);
\bfseries25\mdseries, 355 (1986).

\bibitem{ff} E. Angelopoulos, M. Flato, C. Fr{\o}nsdal, and D. Sternheimer, 
Phys. Rev. D \bfseries 23\mdseries, 1278 (1981). 

\bibitem{gs} G. Scharf, \begin{em}Finite Quantum Electrodynamics\end{em}, 
(Springer, Berlin Heidelberg New York, 1989 and 1995).

\bibitem{br} B. Robertson, 
Phys. Rev. Lett. \bfseries27\mdseries, 1545 (1971).

\bibitem{ae} A. Einstein,
Ann. Phys. \bfseries 49\mdseries, 769 (1916)

\end{thebibliography}
\end{document}